\def\ee#1{\hbox{${} \times 10^{#1}$}}
\def\jj#1#2{\hbox{$J$ = #1$\rightarrow$#2}}
\def\kms{\hbox{km s$^{-1}$}}
\def\samename{{\vrule height0.4pt depth0.0pt width2.0cm \thinspace.}}
\def\ta{{$T_A^*$}}
\def\to{{$\rightarrow$}}
\def\tr{{$T_R$}}
\def\hi{{\ion{H}{1}}}
\def\ci{{[\ion{C}{1}]}}
\def\cii{{[\ion{C}{2}]}}
\def\intu{\hbox{ergs s$^{-1}$ cm$^{-2}$ sr$^{-1}$ Hz$^{-1}$}}
\def\off{{\sc{off}}}
\def\on{{\sc{on}}}
\documentstyle[aaspp]{article}
\begin{document}
\title{Observation of \ci\\
       toward the GL 2591 and W28 A2 Molecular Outflows}
\author{Minho Choi\altaffilmark{1},
        Neal J. Evans II\altaffilmark{2},
        and Daniel T. Jaffe\altaffilmark{3},}
\affil{Department of Astronomy, The University of Texas at Austin,
       Austin, Texas 78712--1083}
\and
\author{Christopher K. Walker\altaffilmark{4}}
\affil{Steward Observatory, University of Arizona,
       Tucson, Arizona 85721}
\altaffiltext{1}{Electronic mail: minho@astro.as.utexas.edu}
\altaffiltext{2}{Electronic mail: nje@astro.as.utexas.edu}
\altaffiltext{3}{Electronic mail: dtj@astro.as.utexas.edu}
\altaffiltext{4}{Electronic mail: cwalker@as.arizona.edu}

\begin{abstract}

Molecular outflows associated with GL 2591 and W28 A2
were observed in the 492 GHz \ci\ line.
Upper limits are set on the \ci\ emission
in the extremely high velocity line wings.
These limits are discussed in terms of 
wind-driven and jet-driven models of molecular outflows.

\end{abstract}

\keywords{ISM: atoms --- ISM: individual (GL 2591, W28 A2) ---
          ISM: jets and outflows --- ISM: kinematics and dynamics}

\section {INTRODUCTION}

Recent observations of molecular outflows in star forming regions have revealed
that, in addition to the usual high-velocity
(\hbox{5 $\lesssim$ $|V-V_0|$ $\lesssim$ 20 \kms}, hereafter HV) outflows,
many sources show extremely high velocity
(\hbox{$|V-V_0|$ $\gtrsim$ 20 \kms}, hereafter EHV) wings on CO lines
(Koo 1989; Masson et al. 1990; Bachiller \& Cernicharo 1990;
Bachiller et al. 1990; Chernin \& Masson 1992; Choi, Evans, \& Jaffe 1993).
The EHV CO emission could arise in the stellar wind,
or it could be ambient molecular gas which has been entrained
(Bachiller \& G\'omez-Gonz\'alez 1992). 

Choi et al. (1993) considered the idea
that the EHV emission arises from CO in the stellar wind.
Their assumption was that the HV CO outflow is driven by the EHV CO outflow
and that the momentum is conserved in this process.
They found that the amount of EHV CO was consistent with this hypothesis
only for low luminosity sources. To explain the deficit of EHV CO in
sources with higher luminosity,
the fraction of C in CO would have to decrease
with source luminosity or rate of production of ionizing photons.
This trend would be plausible
if the missing C were in the form of C$^0$ or C$^+$, as indicated
by models of chemistry in winds (Glassgold, Mamon, \& Huggins 1991).
While EHV wings on lines of C$^+$ would be difficult to detect,
a search for EHV wings from C$^0$ is more tractable (Choi et al. 1993).

Walker et al. (1993) observed the \ci\ fine structure line at 492 GHz
toward about a dozen molecular outflow sources
and showed that the C$^0$ abundance in the HV outflow
is comparable to that in the ambient cloud.
Their velocity coverage and noise level
were not sufficient to detect EHV wings.
If the EHV outflow is not swept-up material,
the C$^0$ abundance in the EHV outflow
is not necessarily the same as in the ambient cloud.
In this paper, we present low-noise observations
of the GL 2591 and W28 A2 molecular outflows in the 492 GHz \ci\ line.
For these two sources,
only $\sim$1\% and  $\sim$5\% of the carbon can be in the form of CO
if the EHV CO wings are caused by gas in the stellar winds (Choi et al. 1993).
If most of the carbon in the wind is in the form of C$^0$,
EHV wings should be readily detectable in these sources.

Recently, outflow models involving collimated jets have been suggested
(Stahler 1993, 1994; Masson \& Chernin 1993; Raga et al. 1993).
In Stahler's picture, gas is entrained in a turbulent mixing process,
producing a continuous range of velocities 
from the ambient gas velocity to velocities near the maximum 
of the HV flows (15--20
\kms), but his model does not yet include specific predictions for higher
velocities.  Masson \& Chernin (1993) address the origin of
discrete EHV features by bow shocks where jets encounter the ambient gas
but do not specifically address smooth extended emission like the
EHV wings. Since Masson \& Chernin invoke a wandering jet to 
produce multiple EHV features, we assume that smooth
extended wings may be the superposition of many such features.
While specific predictions of either CO or \ci\ in EHV wings
are not available for these jet models,
we will argue in \S~4.2 that EHV \ci\ wings would probably not be detectable.

\section {OBSERVATIONS}

We observed GL 2591 on June 20, 1993 and W28 A2 on May 18 and 19, 1993
with the 10.4 m telescope of the Caltech Submillimeter Observatory
(CSO)\footnote{The CSO is operated by the California Institute of Technology
under funding from the National Science Foundation, contract AST 90--15755.}
at Mauna Kea.
We observed at one position in each source given in Table 1,
the peak position of the EHV CO outflow (Choi et al. 1993). 
The data were obtained using a 492 GHz SIS receiver (Walker et al. 1992)
in double-sideband mode with a 500 MHz AOS as the spectrometer.
The antenna temperature (\ta) was calibrated
by the standard chopper-wheel method,
which automatically corrected, to first order,
for the effects of atmospheric attenuation.
The beam size (FWHM) was about 15\arcsec,
and the main beam efficiency ($\eta_{mb}$) was about 0.53 (Phillips 1993).
We present our data in terms of \tr\ = \ta/$\eta_{mb}$.

\begin{table}
{\halign{
#\hfil&\quad\hfil#&\quad\hfil#&\quad\hfil#&\quad\hfil#&\quad\hfil#&\quad\hfil#
&\quad\hfil#\hfil&\quad\hfil#\hfil\cr
\multispan9\hfil TABLE 1 \hfil\cr
\cr
\multispan9\hfil{\sc Source Parameters}\hfil\cr
\cr
\noalign{\hrule \vskip 2pt \hrule \vskip 2pt}
&&&\multispan2\quad\hfil\off1$^a$\hfil&\multispan2\quad\hfil\off2$^a$\hfil
&{\sc Noise}$^b$&{\sc Baseline Intervals}\cr
\noalign{\vskip -9pt}
&&&\multispan2\quad\hrulefill&\multispan2\quad\hrulefill\cr
{\sc Source}&\hfil $\alpha _{1950}$\hfil&\hfil $\delta _{1950}$\hfil
&$\Delta \alpha$&$\Delta \delta$&$\Delta \alpha$&$\Delta \delta$
&(K)&(\kms)\cr
\noalign{\vskip 2pt \hrule \vskip 2pt}
GL 2591&20$^h$27$^m$34\fs9&  40$^\circ$01\arcmin 09\arcsec
       &--120\arcsec &120\arcsec &120\arcsec &--120\arcsec
       &0.062&(--70, --50) and (30, 50)\cr
W28 A2 &17$^h$57$^m$26\fs8&--24$^\circ$03\arcmin 54\arcsec
       &    0\arcsec &120\arcsec &  0\arcsec &--120\arcsec
       &0.41 &(--120, --80) and (80, 100)\cr
\noalign{\vskip 2pt \hrule \vskip 10pt}
\multispan9 \qquad$^a$
Coordinates of \off\ positions relative to the central position.
\hfil\cr
\multispan9 \qquad$^b$
RMS of noise in the \tr\ scale with a velocity resolution of 0.7 \kms.
\hfil\cr}
}
\end{table}

To evaluate the quality of the baseline,
the data were taken by position switching with two \off\ positions
in a symmetric pattern:
\on\to\off1\to\off2\to\on\to\on\to\off2\to\off1\to\on.
The position offsets of the two \off\ positions are shown in Table 1.
To check for contamination in the \off\ positions
and systematic errors in the baseline,
we reconstructed difference spectra of the \off\ positions
by regarding \off1 as the \on\ position
and \off2 as the \off\ position.
Since the difference spectrum
has only half the integration time of the \on\ spectrum,
its RMS noise level is higher than that of the \on\ spectrum
by a factor of about~$\sqrt2$.

The difference spectrum for GL 2591 (Fig. 1 bottom panel)
shows weak emission at the line core and inner blue wing.
In the outer wings, however, the difference spectrum is quite flat.
The difference spectrum for W28 A2 (Fig. 2 bottom panel)
does not show any emission above the noise level.
These spectra indicate that the limit on weak lines is set by the
noise in the \on\ spectrum, rather than by baseline effects.

\begin{figure}[tbp]
\vbox to 3.0cm{\centerline{\it Empty space for Figure 1}}
\caption{
[C {\sc i}] spectrum of GL 2591 with a first order baseline removed
(middle panel).
For comparison, CO and $^{13}$CO $J = 3 \rightarrow 2$ spectra
taken toward the position of the infrared source
are shown in the top panel
(see the map in Choi et~al. 1993 for details).
The dashed spectrum in the middle panel is the [C {\sc i}] line
predicted from the CO $J = 3 \rightarrow 2$ line (see text).
Filled triangles below the baseline of the [C {\sc i}] and CO spectra
are the CO \jj32\ wing boundaries given in Choi et al. (1993).
The bottom panel shows a difference spectrum of the \off1 and \off2 positions.}
\end{figure}

\begin{figure}[tbp]
\vbox to 3.0cm{\centerline{\it Empty space for Figure 2}}
\caption{
[C {\sc i}] spectrum of W28 A2 with a first order baseline removed
(middle panel).
For comparison, CO and $^{13}$CO $J = 3 \rightarrow 2$ spectra
are shown in the top panel.
The dashed spectrum in the middle panel is the [C {\sc i}] line
predicted from the CO $J = 3 \rightarrow 2$ line (see text).
Filled triangles below the baseline of the [C {\sc i}] and CO spectra
are the CO \jj32\ wing boundaries given in Choi et al. (1993).
The bottom panel shows a difference spectrum of the \off1 and \off2 positions.}
\end{figure}

For each spectrum, a first-order baseline was removed.
The last column of Table 1 shows the velocity intervals
considered in the baseline determinations.

\section {RESULTS}

Our main goal is to test the picture
in which a stellar wind at velocities of $\sim$100~\kms\
drives the molecular outflow at $|V-V_0|$ $\lesssim$ 20 \kms\
by sweeping up ambient gas, conserving momentum.
Therefore, we first discuss our results in terms of this picture,
but we also address jet-driven outflows in \S\ 4.2.

The limits on \ci\ in the wind set by our spectra are quite severe.
We have calculated the \ci\ emission in the EHV wing
predicted by the following assumptions:
the mass loss rate is derived from the HV outflow using momentum conservation;
all the missing carbon in the EHV wind
(i.e., carbon not detected by the CO observations)
is in the form of C$^0$ (hence our prediction is an upper limit);
the EHV CO traces the velocity field of the wind;
and the temperature is much higher than 62.5 K,
corresponding to the energy needed to populate the highest level
in the fine structure triplet.
The results are not very sensitive to the last assumption,
as long as the temperature is not very low.

\subsection {GL 2591}

Our \ci\ spectrum of GL 2591 (Fig. 1) agrees with that of Walker et al. (1993)
though their position was about 10\arcsec\ east of ours. 
We failed to detect EHV \ci\ wings, 
even though our noise is about 4 times lower and our velocity coverage
is 10 times wider than that of Walker et al. (1993).
Figure 1 also shows CO lines
taken from about 10\arcsec\ east of the \ci\ observation position.
The line profiles of \ci\ and $^{13}$CO \jj32\ lines agree very well,
which implies that there is no enhancement of atomic carbon abundance
in either the HV or EHV outflows if both lines are optically thin.

The limit on EHV wings is \tr\ $<$ 0.1 K, except near 25 \kms, where there
is a possible (about 3 $\sigma$) feature.
The predictions exceed the detection limit by factors
ranging from 3 to 80 for the blue wing and 8 to 24 for the red wing.

\subsection {W28 A2}

Our \ci\ spectrum of W28 A2 is shown in Figure 2.
This source was not observed by Walker et al. (1993).
As in Figure 1, we also show
the CO and $^{13}$CO \jj32\ lines in the same direction,
as well as the spectrum of the \off\ position.
In this source, the noise is higher, and we set a limit of about 1 K
on the weakest wing we could have detected.
This limit is still much lower than the level of predicted lines.
The blue EHV wing contains a distinct peak between --70 and --50 \kms,
but this peak coincides with a line of CH$_3$OH
($J_{K_{-1}K_1} = 4_{11}\rightarrow 3_{01}$)
at a rest frequency of 492.1607 GHz,
which appears to be prominent in this source, but not in GL 2591.
If we ignore the region of CH$_3$OH emission,
we still set interesting limits on the \ci\ emission.
At the boundaries between inner and outer wings,
the predicted emission is 7 to 9 times our detection limit.

\section {DISCUSSION}

\subsection{Wind-Driven Molecular Outflow Models}

Our failure to detect \ci\ emission in the EHV wings
at levels far below the predictions implies
that some of the assumptions used to make the predictions must be incorrect.
First, we note that the predictions are not very sensitive to temperature,
unless the temperature becomes very low ($<$ 10 K),
and our analysis of the CO EHV emission yielded lower limits to temperatures
which were 16 K for GL 2591 and 94 K for W28 A2,
with the temperatures likely to be considerably higher.
One explanation for not detecting EHV \ci\ emission is
that most of the carbon is ionized,
even though most hydrogen may be in the form of H$^0$.
Some calculations of winds from protostellar disks conclude
that the carbon will be ionized (Safier 1993).
Detection of EHV wings in the \cii\ line, while difficult,
may be possible.
If we make predictions of \cii\ emission
with the same assumptions in the second paragraph of \S\ 3
with C$^+$ instead of C$^0$,
the \cii\ intensities at the boundaries between inner and outer wings
would be 1--4\ee{-11} \intu\ for GL~2591
and 3--4\ee{-11} \intu\ for W28~A2.
Assuming we need to detect a level of 1\ee{-11} \intu\
to clearly demonstrate EHV wings,
this would translate to a \ta\ = 1.0~K,
assuming a filling factor of 0.2,
based on the maps of the EHV wings in CO (Choi et al. 1993),
and an efficiency of 0.6 for a heterodyne instrument on the KAO
(Boreiko, Betz, \& Zmuidzinas 1990).
Using the instrumental parameters described by Boreiko \& Betz (1991),
we calculate an RMS noise of 0.23 K would be reached in 1000 s of
integration with the 20 MHz (3 \kms) filters. 
So far, no observations of these sources in the \cii\ line are available
(Betz 1994).

We have assumed that the carbon in the wind must be
in one of three forms (CO, C$^0$, or C$^+$)
based on models of wind chemistry
that show all other molecules containing carbon to be completely negligible
(Glassgold et al. 1991).
These authors also argue that dust, if it forms at all,
will form outside the region of molecule formation
and will be silicates, rather than carbon grains.
While these chemical models were developed for winds from lower-mass stars,
these basic properties seem quite robust.

Another possible explanation is
that the wind is mostly at velocities
much higher than are traced by the EHV CO,
producing wider and weaker wings.
As discussed by Choi et al. (1993), very wide wings would have been
difficult to detect in CO; they are even harder to detect in \ci.
For GL 2591, our velocity coverage is $\sim$150 \kms\
on each side of the \ci\ line,
so any wing wider than that could not be distinguished from the baseline.
For W28 A2, it is impossible to detect a wider wing
because a CH$_3$OH line on the blue side
and a very bright H$_2$CO line on the red side (outside the velocity range
plotted in Fig.~2) contaminate the baseline.
If we assume a square-shaped wing with a terminal velocity of 300 \kms,
it is possible for an EHV \ci\ outflow to drive the HV CO outflow
with \tr\ = 0.003~K for GL 2591 and 0.04~K for W28 A2.
If present, such low intensity wing emission
would be extremely difficult to detect using available instrumentation.

\subsection{Jet Models}

If sufficiently low limits to the EHV \cii\ emission can be set,
and if the EHV wing traces the velocity field of the wind,
our nondetection of EHV \ci\ would call into question
the standard analysis of mass loss rate from the HV flow,
assuming momentum conservation and a time scale given by the size
of the flow and a characteristic velocity.
These calculations have been challenged on other grounds
(Parker, Padman, \& Scott 1991),
but they also have found some support in detections of \hi\ wings
at the predicted levels (Giovanardi et al. 1992)
and in comparisons of mass loss rates from \ion{Na}{1}
with those deduced from the usual method (Natta \& Giovanardi 1990).

If standard methods have overestimated the mass loss rates,
then jets are an alternative driving source for molecular outflows
(Masson \& Chernin 1993; Raga et al. 1993; Stahler 1994).
Models using jets to drive outflows would be favored
if the carbon in the wind cannot be detected in any form because
the jets have mass-loss rates lower than those inferred
from the standard analysis of wind-driven outflows (Masson \& Chernin 1993). 
Thus, the jet itself would presumably be undetectable in CO or \ci.
However, Hartigan, Morse, \& Raymond (1994) have found evidence
that mass loss rate in jets are higher than previously thought.

In the picture of jet-driven outflows, the
the EHV CO emission would probably arise from ambient gas which has
been entrained by the jet. In the picture of wandering jets, this
material would have been shocked by the jets.
Models of chemistry in shocks
(Neufeld \& Dalgarno 1989) predict ratios of C$^0$ to CO
less than what we inferred for a wind model by factors of $\sim$100.
Therefore the \ci\ intensity predicted by the shocked gas model is
a few times less than our detection limit. If shocks do not occur in the
entrainment process, we can assume that the C$^0$ to CO ratio in the EHV gas is
identical to the column density ratio in the HV gas (0.07 in the blue wing
of GL~2591) observed by Walker et al. (1993). In this case,
the \ci\ intensity prediction is less than our detection limit
by a factor of $\sim$30 for GL~2591.

\section{SUMMARY}

We have set limits on the emission of \ci\ in the EHV wings of
GL~2591 and W28~A2. These limits are considerably below the predictions
of a picture in which the EHV gas is the stellar wind which drives the
HV molecular outflow. Taken at face value, this result would support
the idea that mass loss rates have been overestimated by the standard
analysis of the wind-driven model in these sources, allowing models using
jets to drive the outflows to be viable. However, the wind-driven picture
could still be viable if most of the carbon in the wind is in the form
of C$^+$ or if most of the wind emission is at higher velocities,
producing a weaker, wider line in either CO or \ci. 

\acknowledgements

We thank Constance E. Walker and Yangsheng Wang
for acquiring some of the data.
This work was supported by NSF grants AST-9017710 and AST-9317567,
by a grant from the W. M. Keck Foundation,
and by a David and Lucile Packard Foundation Fellowship.

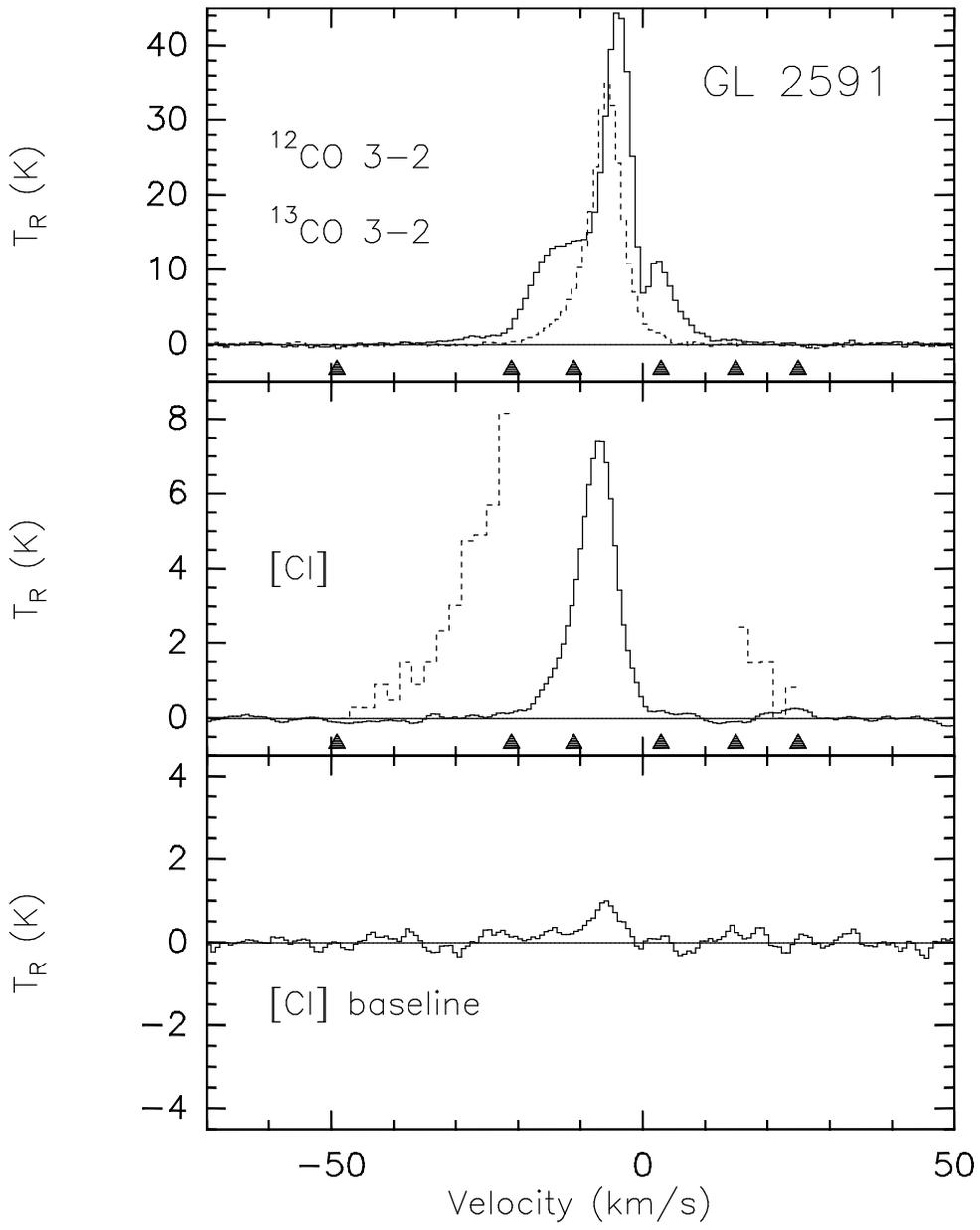
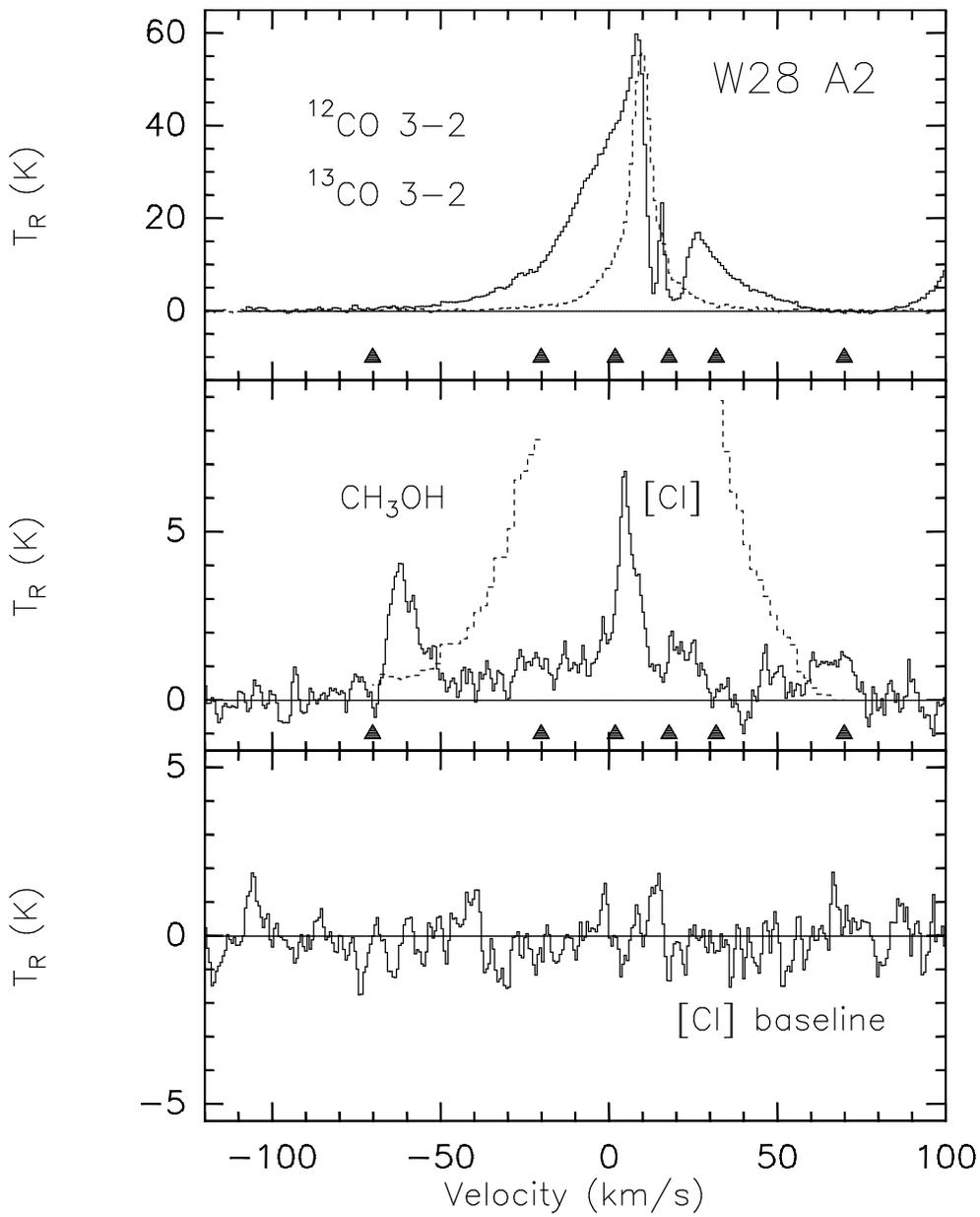
\end{document}